\documentclass[iop]{emulateapj}
\usepackage{multirow}
\usepackage{bm}
\usepackage{amssymb}
\usepackage{amsmath}
\usepackage{latexsym}
\usepackage{graphicx}
\usepackage{ifsym}
\usepackage{bm}
\usepackage{tabularx}

\slugcomment{}
\shortauthors{Wu et al.}

\begin{document}
\title{Magellan AO System $\lowercase{z}'$, $Y_S$, and $L'$ observations of the very wide 650 AU HD 106906 planetary system\footnotemark[$\ast$]}
\footnotetext[$\ast$]{This paper includes data gathered with the 6.5 m \emph{Magellan} Clay Telescope at Las Campanas Observatory, Chile.}

\author{Ya-Lin Wu$^1$, 
 Laird M. Close$^1$, 
 Vanessa P. Bailey$^{2}$,
 Timothy J. Rodigas$^{3,}\footnotemark[5]$,\footnotetext[5]{Hubble Fellow.}  
 Jared R. Males$^{1,}\footnotemark[6]$,\footnotetext[6]{NASA Sagan Fellow.}
 Katie M. Morzinski$^{1}$, 
 Katherine B. Follette$^2$,
 Philip M. Hinz$^1$,
 Alfio Puglisi$^4$,
 Runa Briguglio$^4$, and
 Marco Xompero$^4$}

\affil{$^1$Steward Observatory, University of Arizona, Tucson, AZ 85721, USA; yalinwu@email.arizona.edu\\
$^2$Kavli Institute of Particle Astrophysics and Cosmology, Stanford University, 382 Via Pueblo Mall, Stanford, CA 94305, USA\\
$^3$Department of Terrestrial Magnetism, Carnegie Institute of Washington, 5241 Broad Branch Road, NW, Washington, DC 20015, USA\\
$^4$INAF-Osservatorio Astrofisico di Arcetri, Largo E. Fermi 5, I-50125 Firenze, Italy\\
{\it Accepted for publication in ApJ}}

\begin{abstract} 
We analyze archival data from Bailey and co-workers from the {\it Magellan} adaptive optics system and present the first $0.9~\mu$m detection ($z' = 20.3\pm0.4$~mag; $\Delta z'=13.0\pm0.4$~mag) of the 11~$M_\mathrm{Jup}$ circumbinary planet HD 106906AB b, as well as the 1 and $3.8~\mu$m detections of the debris disk around the binary. The disk has an east--west asymmetry in length and surface brightness, especially at $3.8~\mu$m where the disk appears to be one-sided. The spectral energy distribution of b, when scaled to the $K_S$-band photometry, is consistent with 1800~K atmospheric models without significant dust reddening, unlike some young, very red, low-mass companions such as CT Cha B and 1RXS 1609 B. Therefore, the suggested circumplanetary disk of Kalas and co-workers might not contain much material, or might be closer to face-on. Finally, we suggest that the widest ($a\gtrsim100$ AU) low mass ratio ($M_\mathrm{p}/M_\mathrm{\star} \equiv q\lesssim0.01$) companions may have formed inside protoplanetary disks, but were later scattered by binary/planet interactions. Such a scattering event may have occurred for HD 106906AB b with its central binary star, but definitive proof at this time is elusive.
\end{abstract}


\section{INTRODUCTION}
HD 106906AB b, an 11~$M_\mathrm{Jup}$ planetary mass companion orbiting a 13~Myr \citep{P12} spectroscopic binary ($\gtrsim$2.5~$M_\sun$; \citealt{L16b}) at a projected separation of 650 AU ($7\farcs1$ at 92 pc), was discovered by \cite{Ba14} during the commissioning run of the {\it Magellan} adaptive optics system (MagAO; \citealt{C12}; \citealt{K13}; \citealt{Mo14}). Mounted on the 6.5 m Clay telescope, MagAO utilizes a 585-actuator adaptive secondary mirror to correct for turbulence from 0.6 to 5~\micron. One of MagAO's key science goals is to search for young giant planets and characterize their atmospheres and environments. HD 106906AB b is the first exoplanet discovered by MagAO in a survey targeting stars with unusual debris disks. 

The debris disk around HD 106906AB was first inferred from 24 to 70~\micron~excesses in a {\it Spitzer} survey toward the Scorpius--Centaurus OB association \citep{C05}. It was estimated to have $L_\mathrm{IR}/L_\mathrm{\star} = 1.3\times10^{-3}$ and $M_\mathrm{dust} = 8.2\times10^{-4}~M_\earth$ \citep{C11}. Recently, two teams have successfully imaged this disk with extreme adaptive optics (AO) systems. \cite{L16a} carried out {\it YJHK} imaging with the Spectro-Polarimetric High-contrast Exoplanet Research (SPHERE), while \cite{K15} obtained $H$-band data with the Gemini Planet Imager (GPI). Together with $\beta$ Pic \citep{L09}, Fomalhaut \citep{K08}, HR 8799 \citep{M08}, and HD 95086 \citep{R13}, HD 106906 has joined a family of systems with directly imaged planets and directly imaged debris disks. 

In the GPI and SPHERE images, the disk appears as a nearly edge-on ring (inclination $i\sim$85\degr) with an inner cavity of radius $\sim$50~AU. Particularly interesting is the asymmetry in length and surface brightness (SB) in the $H$ band: the east extension is $\sim$0\farcs1 longer and 10\%--20\% brighter than the west extension. This may imply an asymmetric dust distribution or that the disk is in fact elliptical \citep{K15}. In addition to the bright inner ring, \cite{K15} discovered diffuse nebulosity, including a ``fan'' extending to $\sim$4\arcsec~(370 AU) at a position angle (PA) of $\sim$45\degr~to 90\degr~and a ``needle'' extending to $\sim$6\arcsec~(550 AU) at PA $\sim$285\degr~in archival {\it Hubble Space Telescope (HST)} data. The origin of this intricate morphology, whether sculpted by the interstellar medium and/or the planet, invites future observations.

Also, the formation of HD 106906AB b is hard to understand. At such a wide separation, canonical core accretion is not efficient. \cite{Ba14} favored the in situ star-like formation, although they noted that the observed planet/star mass ratio $M_\mathrm{p}/M_\mathrm{\star}\equiv q<0.01$ is rare among binary stars that form by fragmentation. \cite{L16a} discussed the possibility that the planet might have been captured from somewhere in the cluster. On the other hand, \cite{K15} explored the planet--planet scattering scenario in which the planet initially formed close to the primary and was later scattered by another planet to its current remote orbit. One of their predictions is that in the course of scattering, the plant might acquire material from the protoplanetary disk to form its own circumplanetary disk. Recently, the radial velocity study by \cite{L16b} uncovered the binary nature of the primary, which could also provide a mechanism for scattering the planet outward from an initially tight orbit. The orbital eccentricity is key to constraining the formation and dynamical history of this system, but is hard to measure due to the long orbital period (10,000--100,000 years). Therefore, numerical simulations reproducing the observed disk asymmetry throughout the age of the system (13$\pm$2 Myr; \citealt{P12}) will be invaluable (e.g., \citealt{JPZ15}).

Motivated by the GPI and SPHERE discoveries, we examine archival MagAO data and present the detection of HD 106906AB b at 0.9~\micron, as well as the debris disk at 1 and 3.8~\micron. We quantify the overall SB asymmetry and leave dust grain modeling to future work.

\section{Methodology}
\subsection{Observations}
Using a dichroic dewar window, MagAO can simultaneously operate its two cameras, VisAO (0.6--1~\micron; \citealt{C13}; \citealt{M14}) and Clio2 (1--5~\micron; \citealt{S06}; \citealt{Mo15}). VisAO has a 7.9 mas plate scale and an $8\arcsec \times8\arcsec$ field of view (FOV). On the other hand, the Clio2 camera has a 15.9 mas plate scale and a $5\arcsec\times16\arcsec$ FOV for the specific operating mode in this study.

We imaged the HD 106906 system at $z'$~(0.91~\micron; equivalent width $\Delta\lambda$ = 0.12~\micron) on UT 2013, April 12, and at $Y_S$ (0.98~\micron; $\Delta\lambda$ = 0.09~\micron) and $L'$~(3.8~\micron; $\Delta\lambda$ = 0.68~\micron) on UT 2013, April 4. Weather on both nights was photometric with $\sim$1\arcsec~seeing. We operated the AO system at 300 spatial modes and 1 kHz control loop speed with the primary ({\it V} $\sim$7.8~mag) being the guide star. In total, we obtained 86.16~s $\times$ 10 saturated frames for $z'$, 15~s $\times$ 399/2.273~s $\times$ 34 (saturated/unsaturated) for $Y_S$, and 8~s $\times$ 600/0.164~s $\times$ 60 for $L'$. Data were taken with the instrument rotator fixed in order to allow for angular differential imaging (ADI; \citealt{M06}). These observations spanned parallactic angle rotations of 15\fdg6, 54\fdg7, and 61\fdg9 at $z'$, $Y_S$, and $L'$, respectively. From circular Gaussian fitting to the unsaturated data sets, we estimated the FWHM at $Y_S$ and $L'$ to be 54 mas (6.8 pixels) and 122 mas (7.7 pixels), respectively.

Each of these observation sets was optimized for a different system component: $z'$ was for the planet only; $Y_S$ was only for the disk; and $L'$ was for both the disk and planet. We did not detect the planet at $Y_S$ because we placed the primary around the center of the $8\arcsec \times8\arcsec$ FOV to facilitate disk imaging, causing the planet to lie outside the FOV.

\subsection{$z'$ Reduction and Companion Photometry}
\begin{figure}
\includegraphics[angle=0,width=\columnwidth]{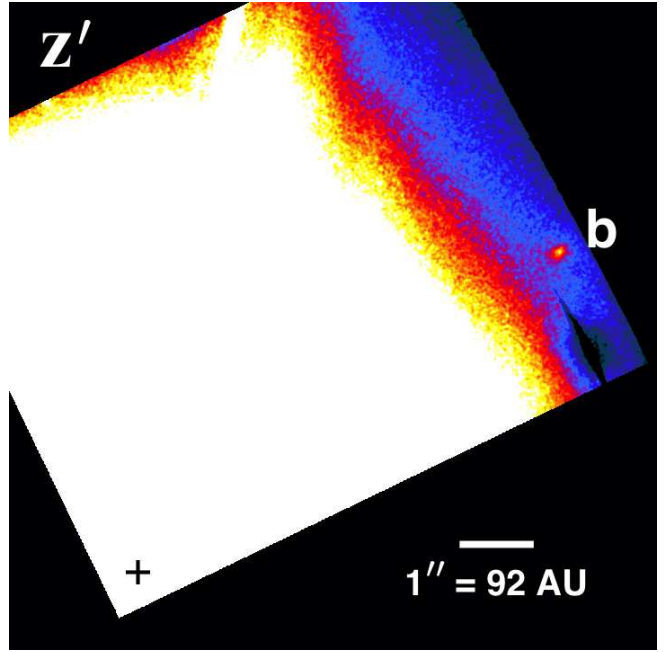}
\caption{HD 106906AB b in $z'$. The cross marks the position of the binary. The dark ``dagger'' below b is an electronic crosstalk. North is up and east is left for Figures \ref{Fig1}--\ref{Fig4}.}
\label{Fig1}
\end{figure}

We carried out $z'$ data reduction using IRAF\footnotemark[7]\footnotetext[7]{IRAF is distributed by the National Optical Astronomy Observatories, which are operated by the Association of Universities for Research in Astronomy, Inc., under cooperative agreement with the National Science Foundation.}(\citealt{T86}, \citeyear{T93}) and MATLAB. After subtracting the dark, we registered images using the Radon transform of diffraction spikes because, in these long-exposure $z'$ data, the star had a highly saturated core of radius $\sim$0\farcs7. The Radon transform projects an image by computing its line integrals along different directions; therefore, the exact positions of diffraction spikes will correspond to the largest integrals. The intersection of spikes then determine the center of a star. We first filtered out the stellar halo using a Gaussian kernel of FWHM $\sim$60~pixels, and masked out unwanted regions such as the saturated core and image boundaries. Next, we performed a Radon transform with MATLAB's built-in function {\it radon} from 0\degr~to 180\degr~in 0\fdg1~increments in order to determine the star's position. After locating the star in these high-pass filtered images to sub-pixel precision ($\sim$0.2 pixels), we shifted and aligned the original non-filtered data accordingly. These registered images were rotated so that north was up and east was left, following the astrometric calibrations in \cite{M14}, and then average-combined (Figure~\ref{Fig1}). For display purposes, we slightly smoothed it with a Gaussian kernel of width = 4 pixels to bring out the companion.

As described in the Appendix, we estimated $z'=20.3\pm0.4$~mag for HD 106906AB b, and a contrast with respect to the primary of $13.0\pm0.4$~mag. The companion's astrometry, $\rho = 7\farcs12\pm0\farcs15$ and PA $= 307\fdg2\pm1\fdg2$, was similar to the literature values. The astrometric error budget includes plate scale uncertainty ($\pm0.012$~mas $\rm{pix}^{-1}$; \citealt{C13}), centroid error, image distortion (\citealt{C13}; \citealt{W15a}), and a $0\fdg3$ rms uncertainty for the MagAO rotator angle (\citealt{C13}; \citealt{M14}).

\subsection{Companion Photometry in the Literature}
We adopted the planet's near-infrared fluxes ($J, K_S, L'$) reported in \cite{Ba14}. For the {\it HST} $F$606$W$ flux, we used the revised measurement in \cite{K15}.

\subsection{$Y_S$ Disk-optimized Reduction}
Dark-subtraction and image registration using cross-correlation were carried out with IRAF. Then, we performed principal component analysis (PCA) using MATLAB's built-in function {\it pca} with 10--30~modes over the entire region $\sim$2\arcsec~from the primary. We did not require a minimum sky rotation between the target frame and the frames used to build the point spread function (PSF). For each PCA mode, we subtracted the reconstructed PSF from the input data, rotated each residual frame to make north up, and average-combined these rotated frames. To bring out the debris disk, we smoothed the combined image with a Gaussian kernel of 1~PSF FWHM (6.8~pixels) derived from the unsaturated data set.

To determine the optimum number of PCA modes, 27 in this case, we followed the procedures delineated in \citeauthor{R12} (\citeyear{R12}, \citeyear{R14}, \citeyear{R15}) to construct a signal to noise per resolution element (SNRE) map for each PCA mode. In short, the SNRE map is the ratio between the original map (smoothed by 1~FWHM) and a noise map. The noise map is constructed by calculating the standard deviation in 1 pixel width concentric annuli in the disk-masked-out smoothed map. The optimum mode generally gives a higher SNRE on both sides of the disk. Figure \ref{Fig2} displays the result processed with 27 modes. The east extension has a median SNRE of $\sim$2.5, and the west extension is detected at SNRE $\sim$1.7.

\subsection{$L'$ Disk-optimized Reduction}
Following the procedures in \citeauthor{Ba13} (\citeyear{Ba13}, \citeyear{Ba14}) and \cite{Mo15}, we used custom MATLAB scripts to linearize counts, flag bad pixels, correct channel bias and electronic crosstalk, subtract the sky background using adjacent nods, and pad and register images. The reader is referred to the Appendix of \cite{Mo15} for a thorough documentation of the detector calibrations. Finally, we performed PCA with 3--15 modes over the central $\sim$2\arcsec. We also did not mask the star, nor did we apply any minimum sky rotation cut to select the frames. In this data set, the disk was severely attenuated above 15 modes. The result with 8 modes is shown in Figure \ref{Fig3}. The median SNRE for the east side is $\sim$2.1, and $\sim$0.7 for the west side. Therefore, we only detected the east extension of the disk at $3.8~\mu$m. This extension is also just visible in Figure 1 of \cite{Ba14}.

For visualization purpose, we also ran PCA with 8 modes for the entire FOV to show the planet and the disk (Figure \ref{Fig4}). 

\subsection{Disk Surface Brightness}
We measured the disk surface brightness in $Y_S$ and $L'$ following \citeauthor{R12} (\citeyear{R12}, \citeyear{R14}, \citeyear{R15}). We rotated the images by 14\degr~clockwise to make the disk horizontal, and centered a square box with width $\sim$1 FWHM (7 pixels for $Y_S$ and 8 pixels for $L'$) on the brightest pixel along each column. We summed over the columns to create the average vertical profile of the disk (perpendicular to the midplane) across that box. We fit a Gaussian to this profile and took the peak as the disk midplane location at the mean stellocentric distance of the box. Next, we centered circular apertures with 1 FWHM in diameter on the fitted midplane and took the median value in each aperture as the disk SB. 

We estimated the background noise as follows. For $Y_S$, we used a sufficiently large number of modes so that the background tends to be Gaussian. Hence, the photometric uncertainties were derived from non-overlapping apertures along azimuthal directions. For $L'$, we did not use enough PCA modes to make the noise exactly Gaussian. The light and dark background ``features'' in Figure \ref{Fig3} were caused by the prevailing wind direction (the AO ``wind butterfly'' effect). We therefore measured the photometric noise only using apertures from similar parts of the sky (i.e., avoiding the dark regions). However, we also calculated the change between the light and dark areas as the background systematic error.

In addition to the non-uniform background, self-subtraction inherent in PCA may also skew our measurements. To understand how it would affect the observed SB, we inserted symmetric artificial disks into the raw data at a PA roughly orthogonal to the real disk and measured the SB ratio after running the reduction pipeline. The disk model was detailed in \cite{R15}, and we adopted the best-fit parameters in \cite{L16a}: 0.6 for the Henyey--Greenstein coefficient $g$, 0\farcs7 for the radius of peak density $r_0$, 85\fdg3 for the inclination $i$, and 10 and $-4.2$ for the inward and outward power-law slopes $\alpha_{\rm in}$ and $\alpha_{\rm out}$, respectively. 

Our artificial disk tests showed that PCA processing effects can lead to spurious asymmetry. The amount and cause of asymmetry is dependent on the number of PCA modes. With fewer PCA modes, the stellar halo is not subtracted uniformly, and uncertainties in the background dominate. With more PCA modes, self-subtraction of the disk dominates. For the numbers of modes used in this paper, we found that PCA processing could induce a spurious asymmetry signal at the $\sim$25\% level. The error budget of the SB profile (shaded areas in Figure \ref{Fig5}, \ref{Fig6}) hence includes the photometric uncertainty, background systematic error (only for $L'$), and this 25\% PCA self-subtraction error.

\begin{figure}
\center
\includegraphics[angle=0,width=\columnwidth]{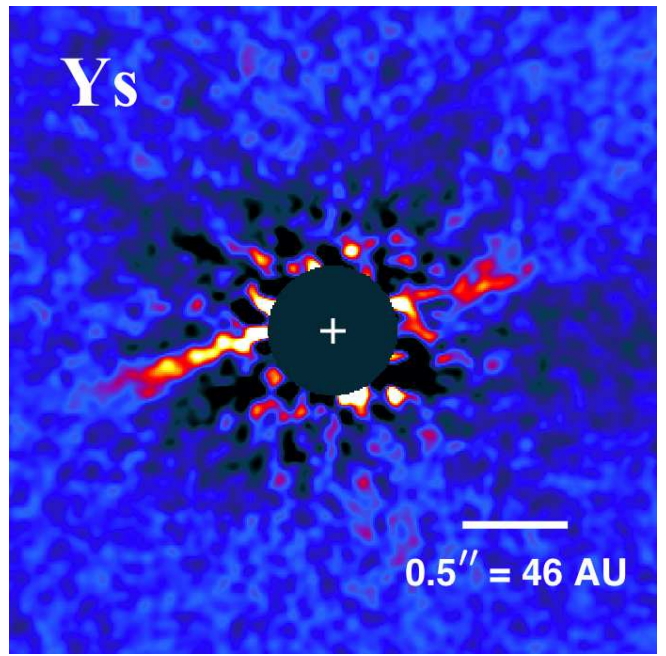}
\caption{Debris disk of HD 106906AB in $Y_S$. We mask out the inner 0\farcs3 for display purpose. The disk is bowed, with an asymmetry in length and brightness.}. 
\label{Fig2}
\end{figure}

\begin{figure}
\center
\includegraphics[angle=0,width=\columnwidth]{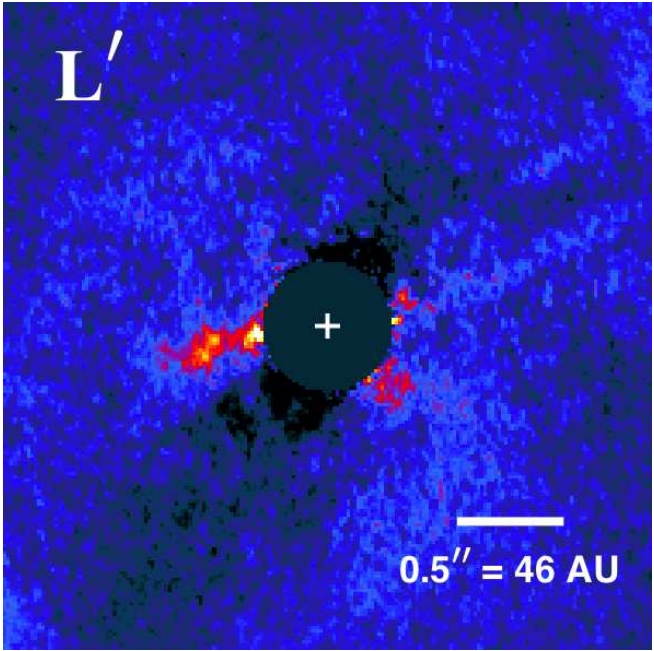}
\caption{Debris disk in $L'$ appears one-sided, tentatively suggesting a higher surface brightness asymmetry at this wavelength.} 
\label{Fig3}
\end{figure}

\begin{figure}
\center
\includegraphics[angle=0,width=\columnwidth]{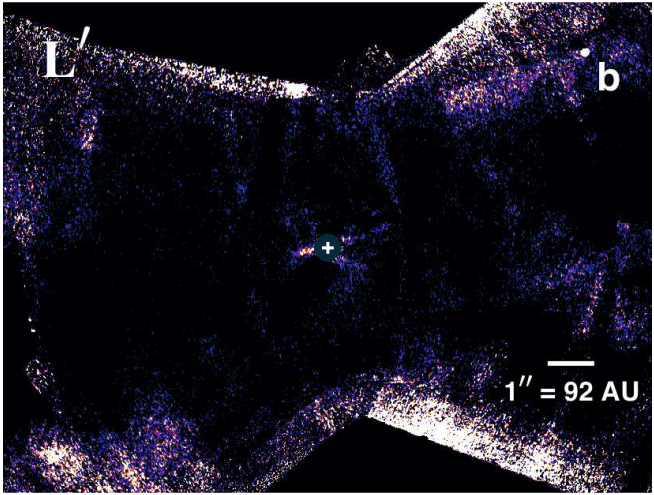}
\caption{HD 106906AB system in $L'$. This image is similar to Figure 1 (processed with classical ADI) in \cite{Ba14}, but with a different contrast to better bring out the disk.} 
\label{Fig4}
\end{figure}

\begin{figure}
\center
\includegraphics[angle=0,width=\columnwidth]{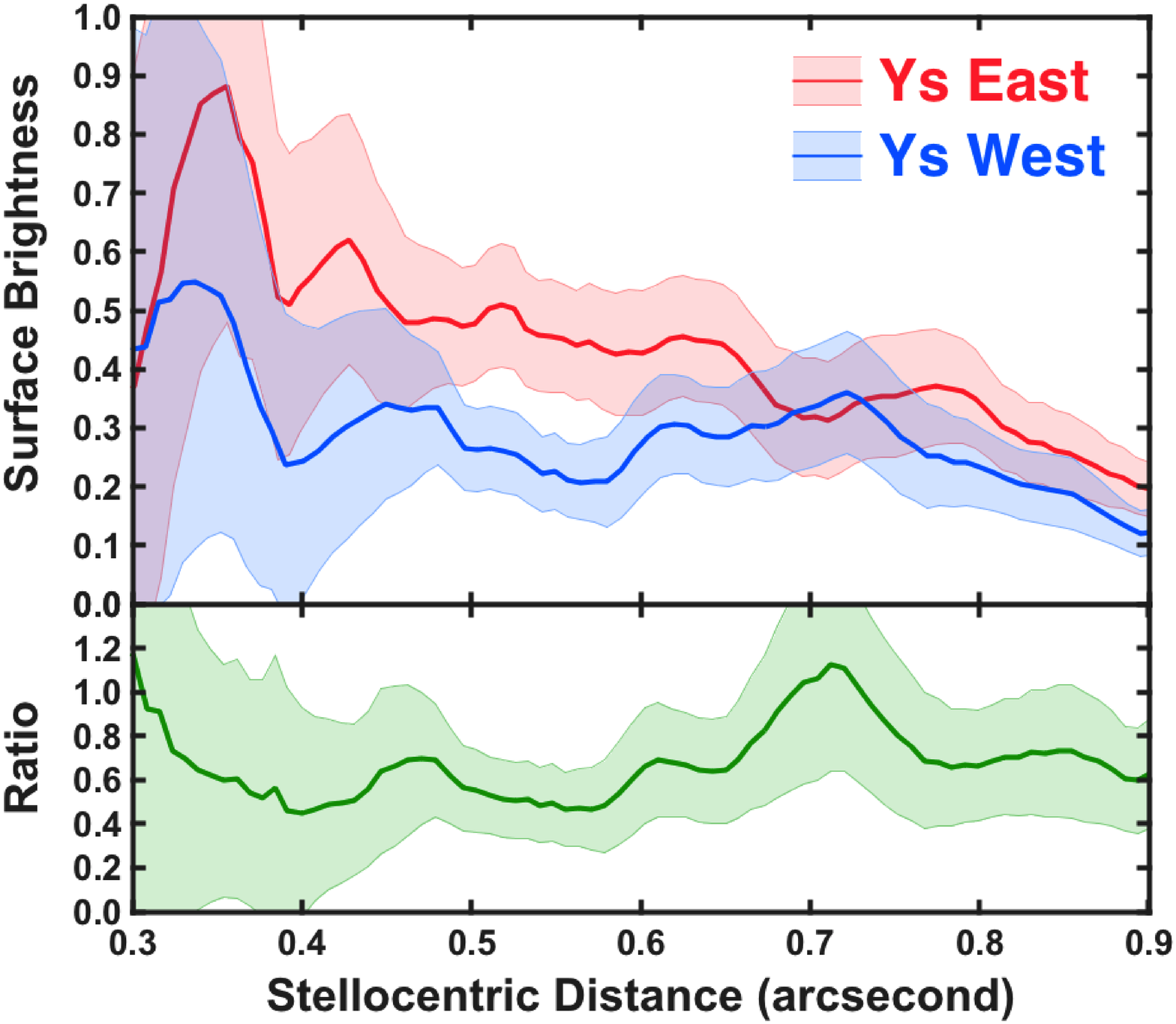}
\caption{SB profile and ratio in $Y_S$. Units for SB are arbitrary. Shaded areas represent the total measurement uncertainties, as described in Section 2.6. Overall, the east side is $\sim$50\% brighter than the west side but, as discussed in the methodology section, part of this asymmetry may come from PCA self-subtraction.} 
\label{Fig5}
\end{figure}

\begin{figure}
\center
\includegraphics[angle=0,width=\columnwidth]{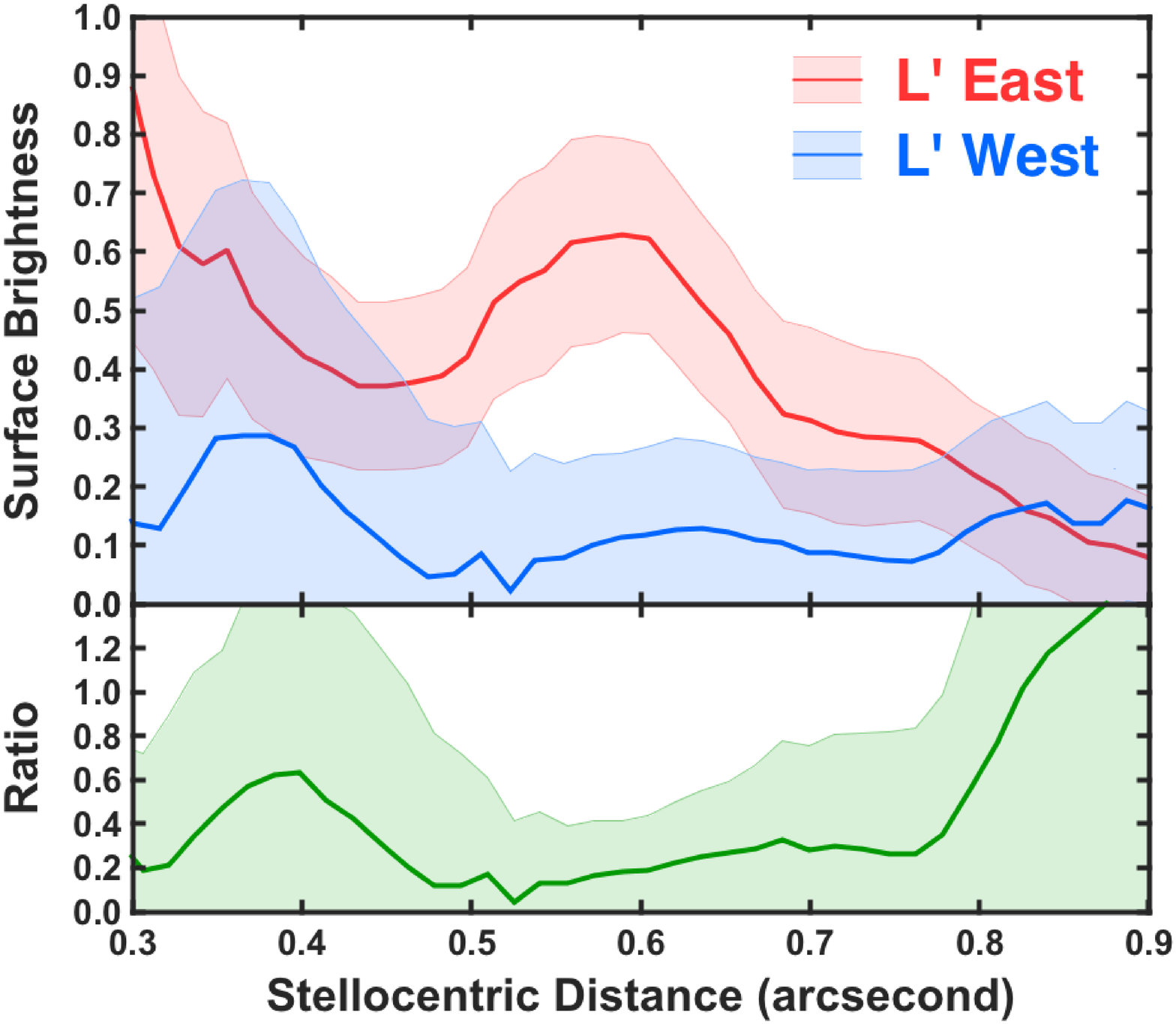}
\caption{SB profile and ratio in $L'$. Since we only detect the east side, the blue curve can be regarded as sky background or an upper limit of the west disk. Overall, the east side is about 1.5--5 times brighter than the upper limit on the west side, albeit with significant uncertainties. As a result, the SB asymmetry in $L'$ is probably higher than $Y_S$ as well as other wavelengths observed by GPI and SPHERE.}
\label{Fig6}
\end{figure}

\begin{figure}
\center
\includegraphics[angle=0,width=\columnwidth]{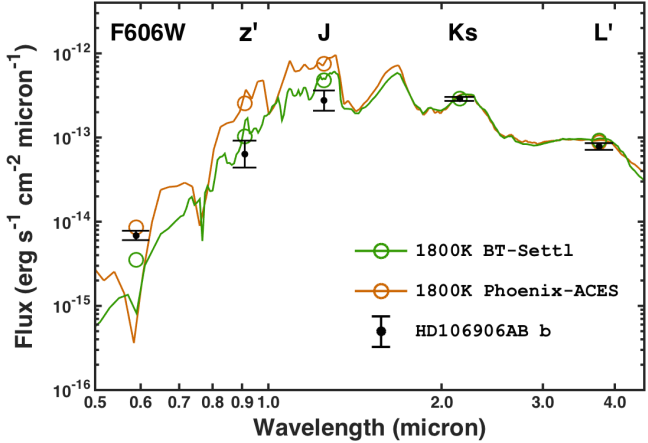}
\caption{SED of HD 106906AB b compared with the 1800~K atmospheric models. We normalize models at $K_S$. Squares represent synthetic fluxes. Unlike CT Cha B and 1RXS 1609 B (\citeauthor{W15a} \citeyear{W15a}, \citeyear{W15b}), HD 106906AB b does not have significant dust reddening.}
\label{Fig7}
\end{figure}

\begin{figure}
\center
\includegraphics[angle=0,width=\columnwidth]{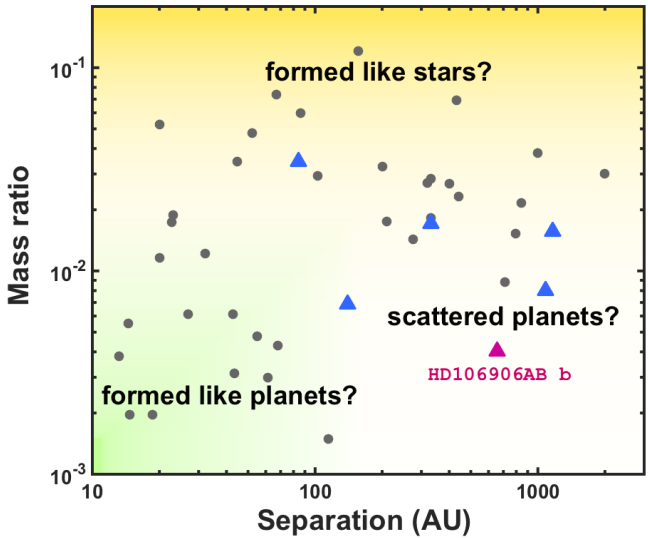}
\caption{Mass ratio versus projected separation for directly imaged companions from 10 to 3000 AU. We exclude companions to highly evolved stars (like white dwarfs) or very low-mass primaries (brown dwarfs). For known circumbinary systems (marked by colored triangles), stellar mass is taken to be the sum of both primary components. We label three formation scenarios for three different areas of parameter space (see text). Other than Fomalhaut b, HD 106906AB b has the lowest mass ratio beyond 100 AU. We note that if HD 106906AB b is coplanar with the debris disk, then its physical separation can be as large as $\sim$3000 AU (\citealt{K15}; \citealt{L16a}). Data are retrieved from the exoplanets.eu database (and LkCa 15 b, c from \citealt{S15}).}
\label{Fig8}
\end{figure}

\section{Results}
\subsection{Debris Disk}
Figure \ref{Fig2}--\ref{Fig4} display MagAO imaging of the debris disk. At $1~\mu$m, both sides of the disk are bowed toward the star, which is similar to the earlier GPI and SPHERE images. The bow-shaped structure is similar to what would be expected for a nearly edge-on disk, e.g., HD 32297 (\citealt{S05}, \citeyear{S14}; \citealt{R14}) or HD 15115 (\citealt{K07}; \citealt{R12}; \citealt{Mz14}; \citealt{S14}). On the other hand, at $3.8~\mu$m, the disk looks one-sided; only the east part is detected. 

In Figure \ref{Fig5} and \ref{Fig6}, we plot the disk SB profile and the ratio between 0\farcs3 and 0\farcs9. At $1~\mu$m, the eastern disk is $\sim$50\% brighter than the western counterpart at $\sim$$1\sigma$~level. However, as described previously, PCA self-subtraction can easily create a $\sim$25\% asymmetry. Thus, the true SB asymmetry in $Y_S$ may be around 20\%, which is similar to what GPI and SPHERE have found. On the contrary, at $3.8~\mu$m, the east side is likely $>$50\% brighter than the other side, even after taking self-subtraction into account. Hence, we hypothesize a wavelength-dependent asymmetry, perhaps arising from different surface densities or grain properties. Detailed dust grain modeling is essential to test this idea; however, it is beyond the scope of this paper.

\subsection{Testing for a circumplanetary disk with SED fitting}
Judging from the planet's infrared color and marginally wider Airy ring, along with the unusual debris disk morphology, \cite{K15} proposed that HD 106906AB b might have captured material from the disk in past dynamical interactions, thereby forming a circumplanetary disk. They also suggested that the optical {\it HST} flux of HD 106906AB b is greater than theoretical predictions, so that this excess may come from stellar light scattered by the circumplanetary disk. We investigate this hypothesis with the addition of our $z'$ photometric point.

Figure \ref{Fig7} compares the planet's spectral energy distribution (SED) to 1800~K atmospheric models (Phoenix-ACES, \citealt{B11a}, \citeyear{B11b}; BT-Settl, \citealt{A11}). We normalize the models in the $K_S$ band because it has a higher signal-to-noise ratio (S/N) compared to other filters and is less affected by extinction than $J$. We also note that normalizing at $L'$ (our other high S/N photometric point) produces a result very similar to $K_S$. 

We look for the presence of three potential disk signatures in the SED: (1) reddening at $F$606$W-J$, (2) thermal emission from hot dust at $K_S$ and $L'$, and (3) scattered light at $F$606$W$. Although the Phoenix model tends to overestimate the $z'$ and $J$ fluxes, both models give overall reasonable fits, consistent with $T_\mathrm{eff}=1800\pm100~\mathrm{K}$ derived by \cite{Ba14} using evolutionary models. This suggests that HD 106906AB b does not have much, if any, dust extinction, in contrast to $A_V\sim$ 3--5 mag for CT Cha B and 1RXS 1609 B in our previous studies of very red planetary mass objects (\citealt{W15a}, \citeyear{W15b}). Furthermore, this implies that HD 106906AB b is not a more massive brown dwarf which is reddened to appear like a cool, low-mass object. We also note that the $L'$ flux appears purely photospheric, with no excess from hot circumplanetary dust emission. As with the potential reddening at $z'$ and $J$, the potential scattered light excess at $F$606$W$ is model-dependent, with the $F$606$W$ photometric data falling between the two models.

Of course, we caution that the above interpretations hinge on the chosen normalization reference. If the models were scaled to $J$ band as in \cite{K15}, then the models would underestimate the infrared fluxes and imply a $J-K_S$ excess, as posited in that paper. However, we note that even with $J$ normalization, there is no significant $K_S-L'$ excess color above the 1800~K photosphere (see Figure 3 of \citealt{Ba14}), while some $K_S-L'$ excess would be expected if the $J-K_S$ color was due to thermal emission from a hot disk. We prefer $K_S$ (or $L'$) normalization over $J$ due to the much higher S/N detection of b at $K_S$ (or $L'$).

Clearly, future multi-wavelength observations are critical to test the existence of this circumplanetary disk. New or improved $J$- and $H$-band measurements would be very helpful to constrain the planet's SED, while (sub)millimeter interferometry would be sensitive to the parent body population in the disk. At the moment, we can only speculate that the proposed circumplanetary disk may be rather depleted and not able to sustain on-going accretion, or it is in a nearly face-on configuration so that along our line of sight is optically thin with no extinction.

\section{Discussion}
In Figure \ref{Fig8}, we plot the planet-to-star mass ratio $q$ versus the projected separation for directly imaged companions. Objects close to their parent stars ($\lesssim$100 AU) can form inside a protoplanetary disk, e.g., LkCa 15 b, c \citep{S15}, but this planet-like formation (core accretion) is inefficient in situ for wide companions. Though suffering from small number statistics and observational bias, in Figure \ref{Fig8} most of the wide companions beyond 100 AU have $q \gtrsim0.01$, while HD 106906AB b has the lowest ratio $q \sim0.004$. If HD 106906AB b and most of the wide companions follow in situ binary-star-like fragmentation formation, then this may imply that binary formation by fragmentation is very inefficient, if not ceased, at mass ratios $q \lesssim0.01$ (e.g., \citealt{LL76}; \citealt{ZS09}). 

However, Figure \ref{Fig8} also shows that circumbinary companions (colored triangles) have large separations around 100--1000 AU; thus, three-body interaction may play a role in scattering this subset of objects to ultra-wide orbits (e.g., \citealt{RC01}). This is appealing for the HD 106906 system as \cite{L16b} recently discovered that HD 106906 is a spectroscopic binary. Therefore, it is possible that HD 106906AB b was initially born in a massive protoplanetary disk, and then scattered out as the least-massive member from a close encounter with one of the stars to its current position. In this manner, we can explain how such a wide 650 AU solar system could have formed. 

Summarizing the above conjectures, in Figure \ref{Fig8}, we label three regimes. Objects with high mass ratios ($q\gtrsim0.01$) form like stars and brown dwarfs. Objects with low mass ratios ($q\lesssim0.01$) and close to their parent stars are born in protoplanetary disks, such as the HR 8799 and LkCa 15 planets. Objects with low mass ratios but at ultra-wide separations may either form like stars or via scattering events. A spectroscopic survey may provide clues to the relative abundance of spectroscopic binary primaries of wide very low-mass companions. Hence, if spectroscopic binaries like HD 106906AB are common, then scattering is very likely the cause of the widest planetary systems. 

\section{Summary}
We present the MagAO $z'$ detection of the circumbinary planet HD 106906AB b, as well as the $Y_S$ and $L'$ detections of the primary stars' debris disk. The disk seems to have a higher surface brightness asymmetry in $L'$ than $Y_S$. The SED of the planet, when scaled to the $K_S$ (or $L'$) flux, suggests that it does not harbor a warm, dusty circumplanetary disk, or the disk is closer to face-on. Finally, we suspect that circumbinary planets at wide separations ($\gtrsim$100 AU) like HD 106906AB b may be scattered to their current remote orbits by interactions with one of their primary stars.

\acknowledgements
We thank the anonymous referee for helpful comments. We are grateful to the MagAO development team and the {\it Magellan} Observatory staff for their support. Y.-L.W. and L.M.C. are supported by the NASA Origins of Solar Systems award. J.R.M. and K.M.M. were supported under contract with the California Institute of Technology (Caltech) funded by NASA through the Sagan Fellowship Program.

\appendix
\section{Photometry on the 0.9~$\mu\lowercase{m}$ detection of the planet}
Our $z'$ photometry suffers from some complications. First, we could not obtain unsaturated images of the primary to perform relative photometry. Therefore, for photometric calibration, we used $z'$ data from two other stars observed during the same observing run. Second, anisoplanatism (the decrease in AO quality with increasing separation from the guide star) causes PSF variability across the FOV. We must correct for this variability when using on-axis standard stars as photometric calibrators for our off-axis planet. In addition to anisoplanatic effects, intrinsic variability in the quality of AO correction between HD 106906AB and these photometric standards also contributes to our uncertainty in photometric calibration. Here, we describe our efforts to better measure the planet's flux by fitting the effects of anisoplanatism on the planet and applying them to unsaturated PSF stars of known $z'$ magnitudes.

\begin{figure}[h]
\centering
\includegraphics[angle=0,width=0.33\columnwidth]{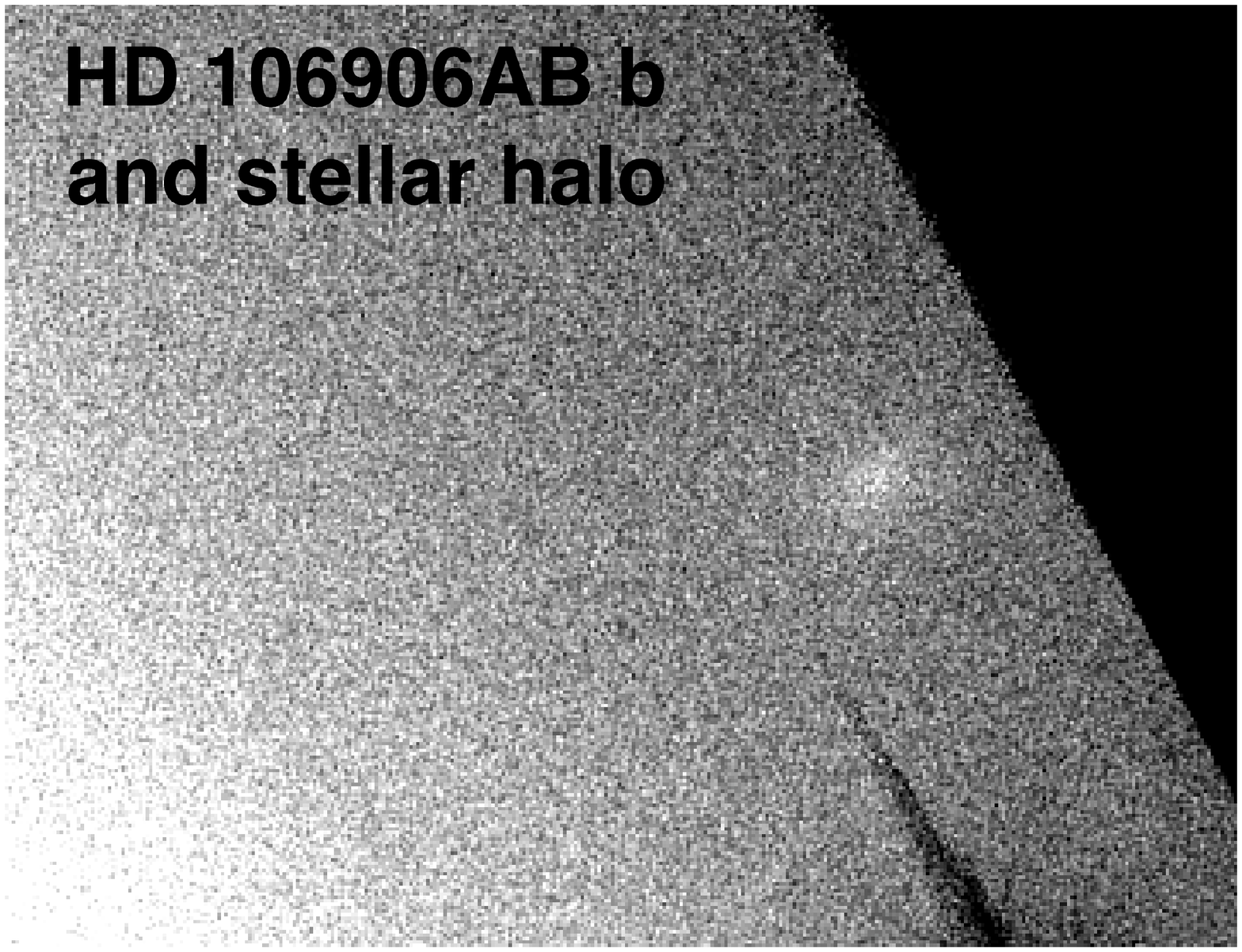}
\includegraphics[angle=0,width=0.33\columnwidth]{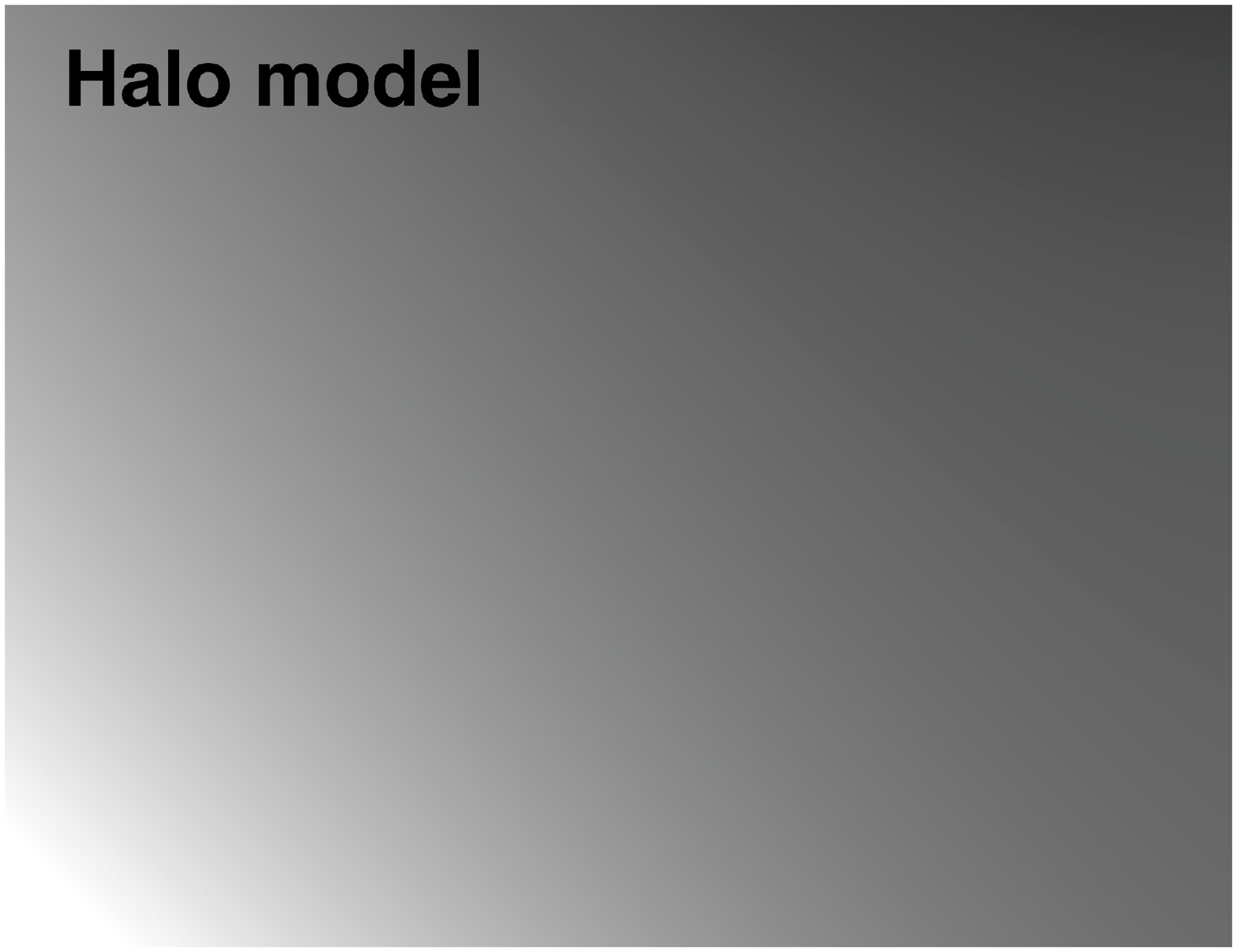}
\includegraphics[angle=0,width=0.33\columnwidth]{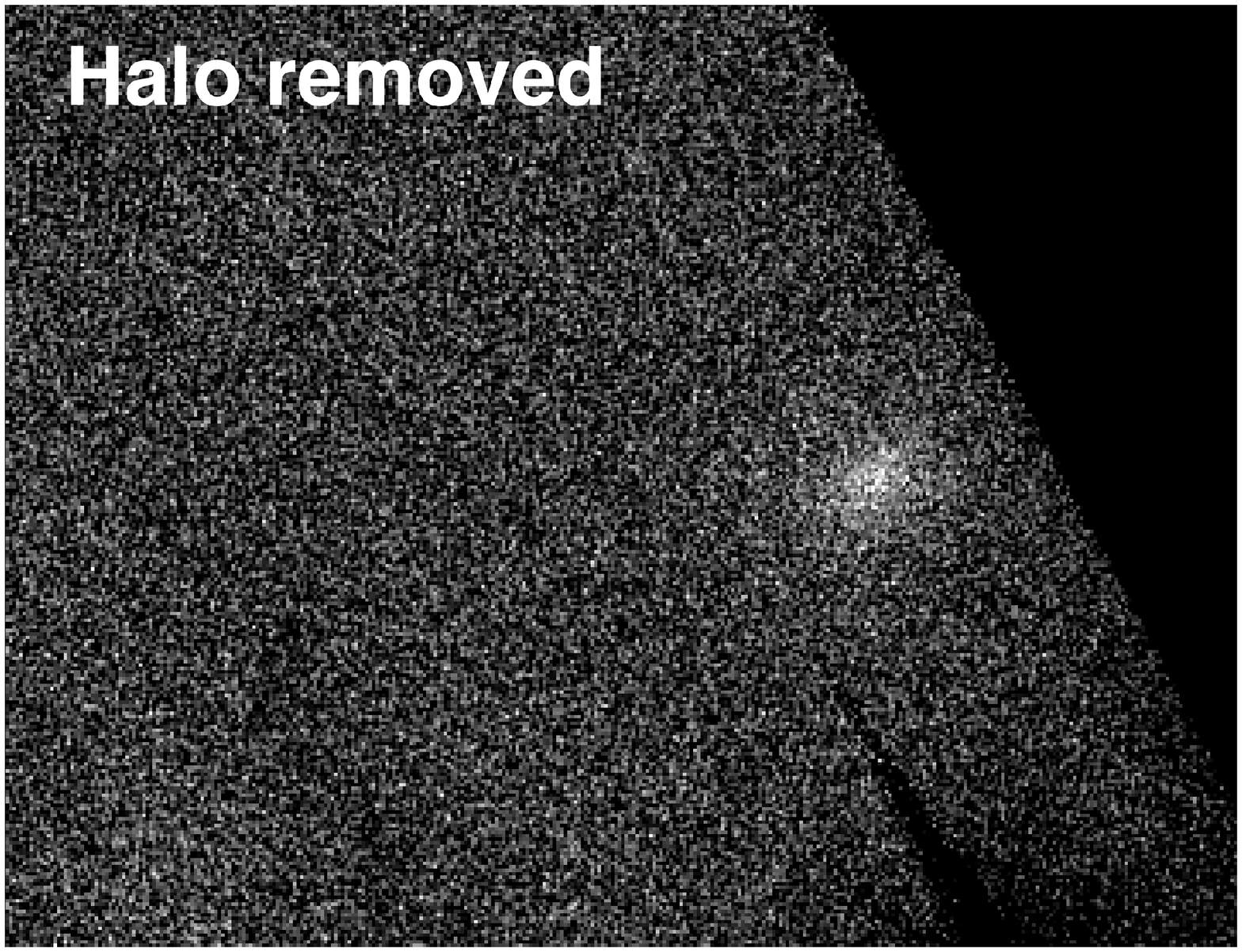}
\caption{Left: the $z'$ image has a strong intensity gradient due to the stellar halo. Middle: we fit a third-order polynomial to the background. Right: subtracting this halo model from the original image, we obtained a uniform background.}
\label{FigA1}
\end{figure}

\begin{figure}[h]
\centering
\includegraphics[angle=0,width=0.49\columnwidth]{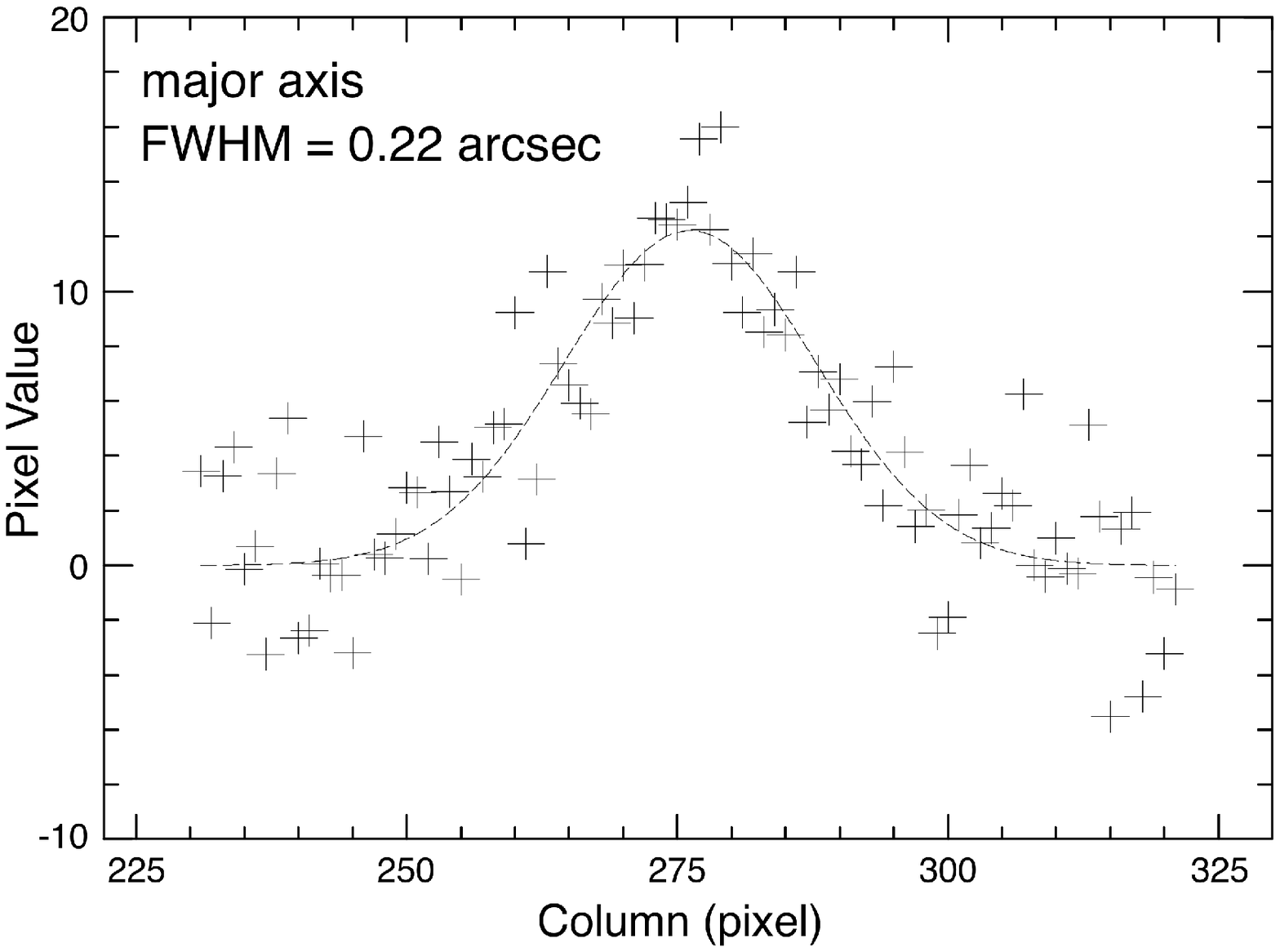}
\includegraphics[angle=0,width=0.49\columnwidth]{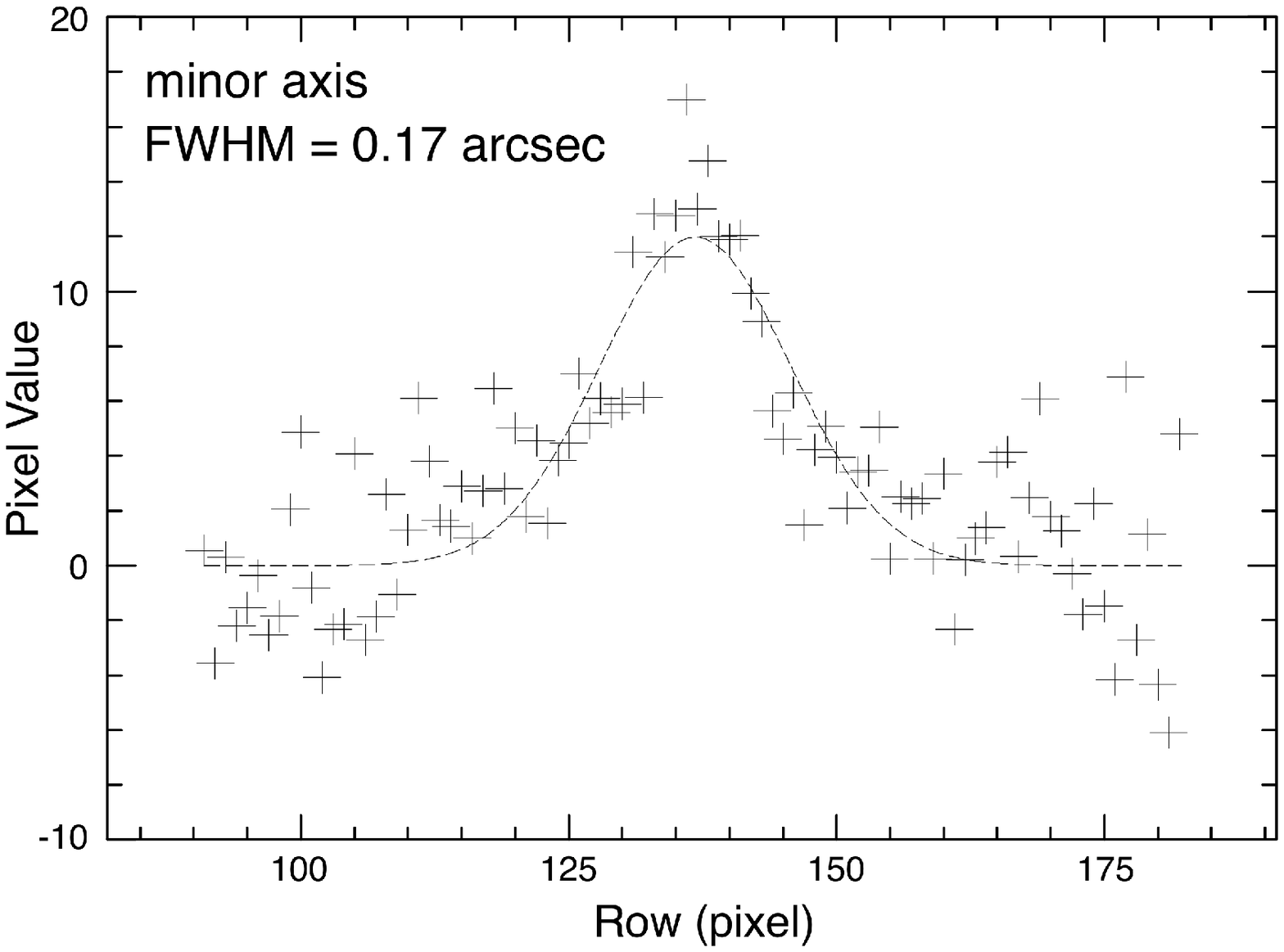}
\caption{The planet has a $0\farcs22\times0\farcs17$ elliptical PSF due to anisoplanatism. Note that the VisAO camera has a 7.9 mas plate scale.}
\label{FigA2}
\end{figure}

We modeled the PSF of the planet with an elliptical Gaussian after removing the contribution from the primary stars' halo. As the entire FOV was immersed in the halo, we fit the background around HD 106906AB b using a third-order polynomial and subtracted it from the image (Figure \ref{FigA1}). Then, we rotated the resulting image by 37\degr~clockwise so that the major axis of the PSF was horizontal. After fitting a Gaussian to each axis independently, we found 0\farcs22 and 0\farcs17 for the major and minor axes, respectively (Figure \ref{FigA2}). We adopted the geometric mean of both axes, 0\farcs19, as the effective FWHM.

\begin{figure}[h]
\centering
\includegraphics[angle=0,width=0.33\columnwidth]{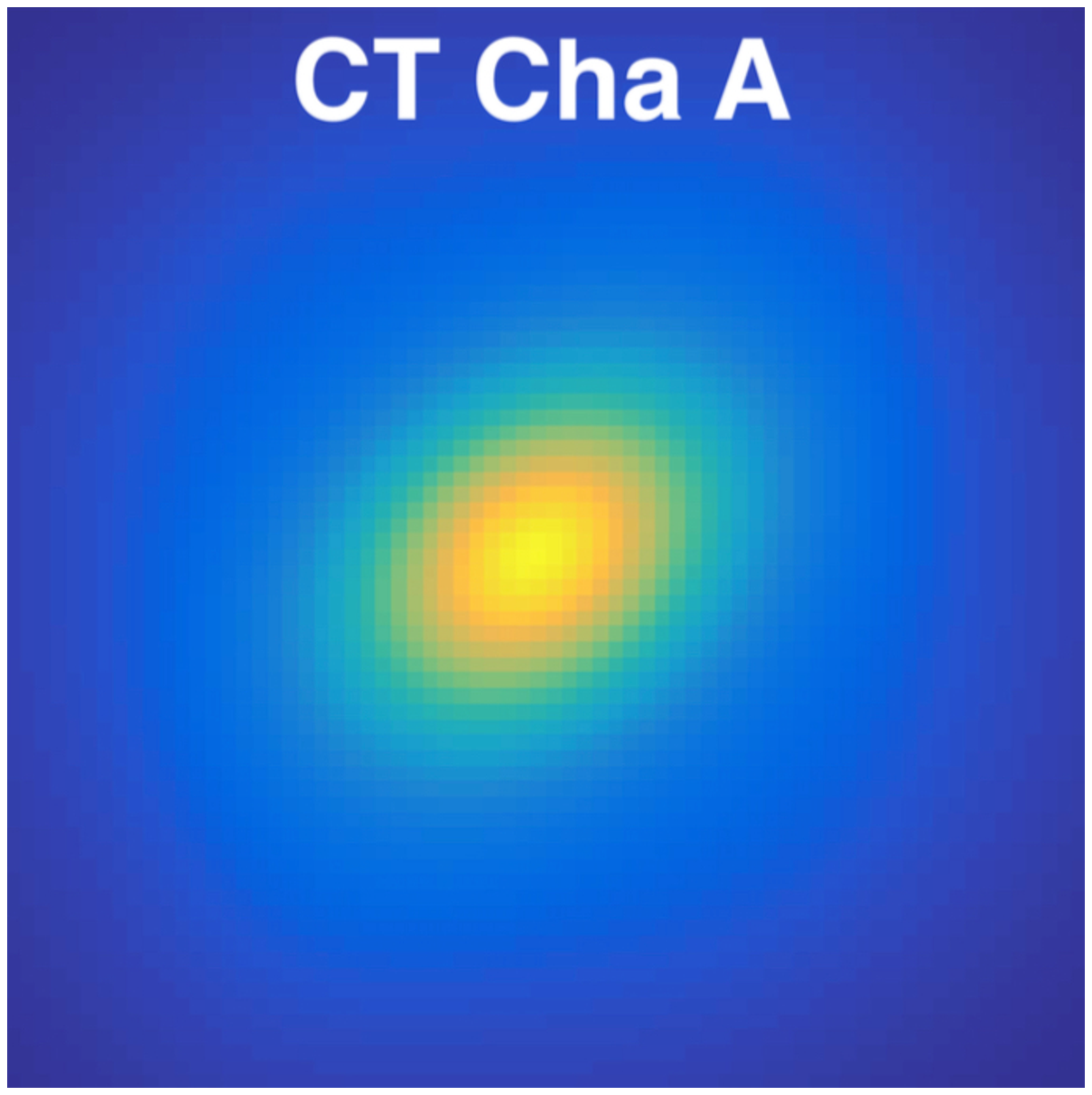}
\includegraphics[angle=0,width=0.33\columnwidth]{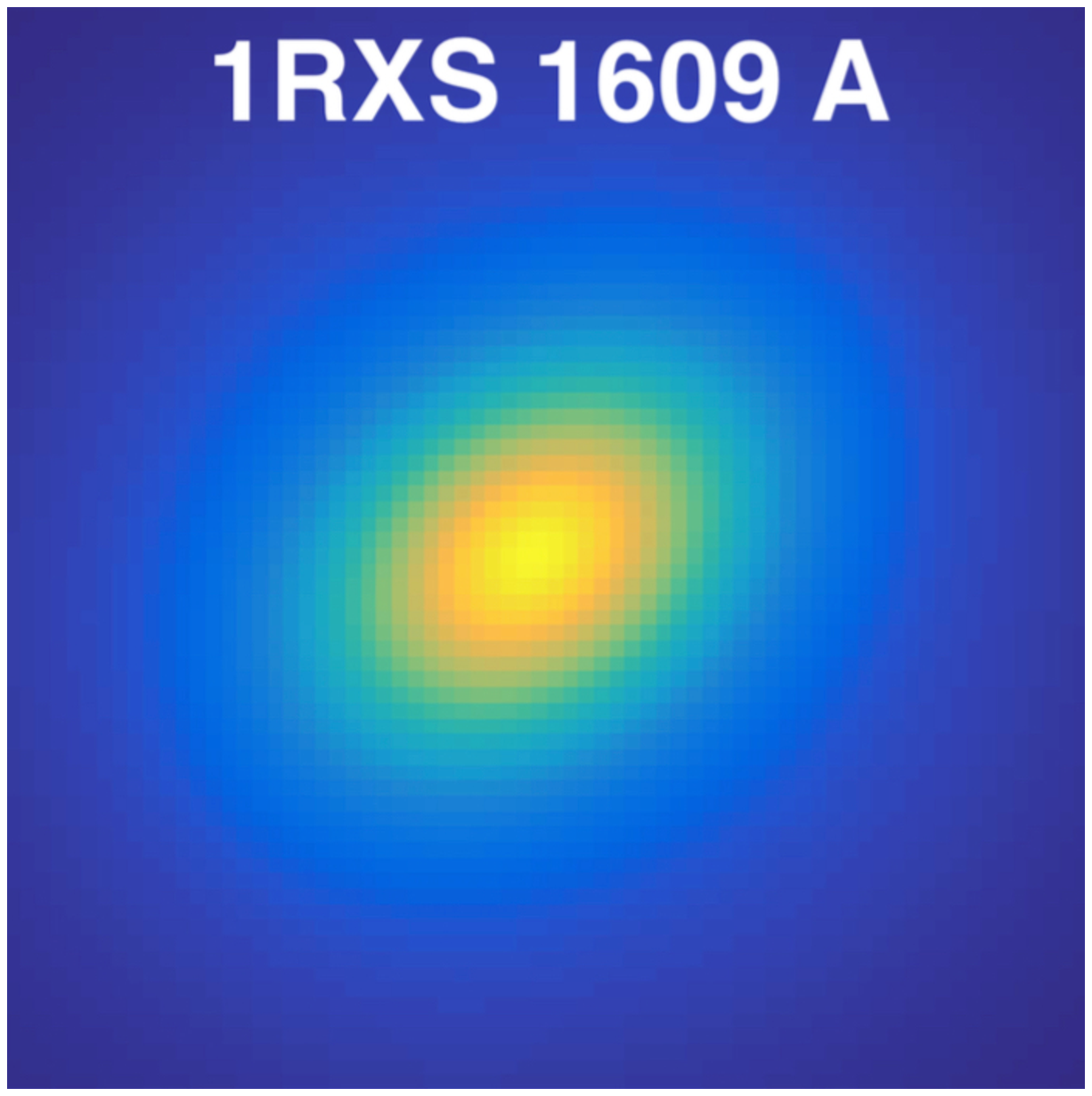}
\includegraphics[angle=0,width=0.33\columnwidth]{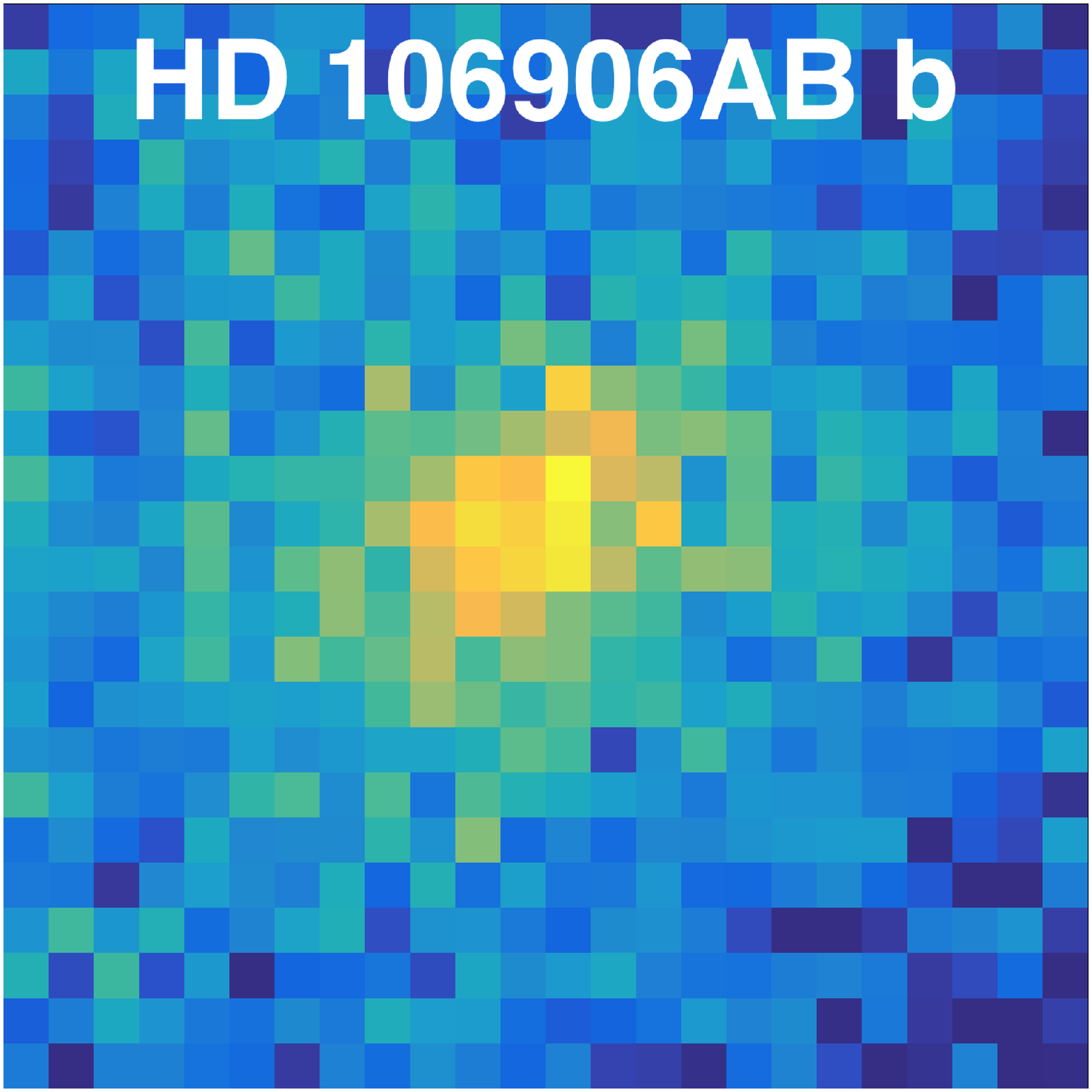}
\caption{We artificially applied the anisoplanatic effects to CT Cha A and 1RXS 1609 A by convolving their PSFs with elliptical Gaussian functions so that their major and minor axes matched that of HD 106906AB b. To facilitate visual comparison, in the right panel we applied a 3 by 3 block average so that the planet's PSF elongation can be readily seen.}
\label{FigA3}
\end{figure}

We next compared the planet to CT Cha A ($z'=10.50\pm0.05$~mag; \citealt{W15a}) and 1RXS 1609 A ($z'=10.60\pm0.06$~mag; \citealt{W15b}). Both stars were observed on UT 2013, April 6. Weather was stable and photometric with $\sim$0\farcs7~seeing. To approximate the Strehl ratio degradation seen in the planet's PSF, we convolved both stars with elliptical Gaussians so that they shared a similar ellipticity with the planet (Figure \ref{FigA3}). In the absence of detailed modeling of anisoplanatic effects, this procedure broadly compensates for the systematic errors in relative photometry between our off-axis target and on-axis standards.

We performed aperture photometry with 0.5, 1.0, 1.5, and 2.0 effective FWHMs (0\farcs095, 0\farcs19, 0\farcs285, 0\farcs38) in diameter. We found that the contrast between the planet and reference stars remained stable with respect to aperture sizes. For CT Cha A, the contrasts were, respectively, 9.66, 9.64, 9.61, and 9.62 mag, suggesting $z'\sim20.1$ mag for the planet. For 1RXS 1609 A, the contrasts were, respectively, 9.80, 9.82, 9.79, and 9.77 mag, suggesting $z'\sim20.4$ mag. There was likely a 0.3 mag systematic error between these two reference stars, possibly arising from slightly different AO correction qualities (CT Cha A had a lower Strehl ratio than 1RXS 1609 A). On the other hand, air mass effect is not likely to significantly contribute to the photometric uncertainty. Detailed atmospheric models show that over our range of air masses ($\sim$1.2 for HD 106906AB b; $\sim$1.0 for 1RXS 1609 A; $\sim$1.5 for CT Cha A), there is less than a 5\% difference theoretically predicted by integrating telluric models over those different air masses at $z'$ \citep{W15a}. Hence, we derive $z' = 20.3\pm0.4$ mag for HD 106906AB b from taking the average of both measurements. The total error budget includes the 0.3 mag systematic error, as well as uncertainties in relative and absolute photometry.

Finally, inferring the primary's $z'$ brightness by interpolation between its $BVJHK$ magnitudes (from the SIMBAD data base), we calculate a $z'$ contrast of $13.0\pm0.4$~mag between the spectroscopic binary and HD 106906AB b.\\
%\clearpage

\end{document}